# Large current modulation in exfoliated-graphene/MoS$_2$/metal vertical heterostructures

Rai Moriya[1,a)], Takehiro Yamaguchi[1], Yoshihisa Inoue[1], Sei Morikawa[1], Yohta Sata[1], Satoru Masubuchi[1,2], and Tomoki Machida[1,2,b)]

[1] *Institute of Industrial Science, University of Tokyo, 4-6-1 Komaba, Meguro, Tokyo 153-8505, Japan*

[2] *Institute for Nano Quantum Information Electronics, University of Tokyo, 4-6-1 Komaba, Meguro, Tokyo 153-8505, Japan*

Graphene-based vertical field effect transistors have attracted considerable attention in the light of realizing high-speed switching devices; however, the functionality of such devices has been limited by either their small ON-OFF current ratios or ON current densities. We fabricate a graphene/MoS$_2$/metal vertical heterostructure by using mechanical exfoliation and dry transfer of graphene and MoS$_2$ layers. The van der Waals interface between graphene and MoS$_2$ exhibits a Schottky barrier, thus enabling the possibility of well-defined current rectification. The height of the Schottky barrier can be strongly modulated by an external gate electric field owing to the small density of states of graphene. We obtain large current modulation exceeding $10^5$ simultaneously with a large current density of ~$10^4$ A/cm$^2$, thereby demonstrating the superior performance of the exfoliated-graphene/MoS$_2$/metal vertical field effect transistor.

a) E-mail: moriyar@iis.u-tokyo.ac.jp
b) E-mail: tmachida@iis.u-tokyo.ac.jp

The discovery of graphene (Gr), which is a one-atom-layer-thick carbon sheet, has attracted considerable attention in condensed matter research into two-dimensional (2D) crystals and van der Waals heterostructures [1,2]. These 2D crystals have layered structures in the bulk form, and each layer has strong in-plane bonding. On the other hand, interlayer coupling is sustained by van der Waals forces, thus enabling the mechanical exfoliation of these crystals down to one monolayer. Further, the fabrication of van der Waals heterostructures based on these 2D crystals has recently been demonstrated [1]. The combination of atomic-layer-level control of 2D crystal material and the technological capability of single-crystal heterostructure fabrication using the van der Waals force, which is, in principle, free from lattice mismatch and inter diffusion at the interface, can potentially lead to a new era in materials science. The most unique properties of such heterostructures concern vertical current transport across the van der Waals interface. Recently, the realization and demonstration of such a vertically stacked heterostructure has revealed its potential high-performance application in tunneling transistors, photovoltaic cells, and flexible devices [3,4,5,6,7,8,9]. A large ON-OFF ratio of $10^5$–$10^6$ has been demonstrated in Gr/$WS_2$/Gr vertical tunnel transistors and Gr/Si barristors. However, in these devices, the ON current density tends to be small (around $10^1$–$10^2$ A/cm$^2$). In comparison with these devices, the Gr/$MoS_2$/metal (GMM) vertical heterostructure provides significant advantages in terms of a large driving current and improved quantum efficiency [6,7]. In particular, the GMM device exhibits an ON current density of ~$10^3$ A/cm$^2$, which is a significant improvement over the ON current density values of other Gr-based vertical transistors. These advantages are strongly related to the presence of the metal electrode on one side of the structure. However, thus far, the experimentally observed ON-OFF ratios in such GMM structures has been limited to values below ~$10^3$ [6,10]. This is smaller than those achieved using Gr/$WS_2$/Gr and Gr/Si devices, although all these devices rely on the same current modulation mechanism (the gate modulation of graphene's Fermi level). It has been considered that this reduction in the ON-OFF ratio in GMM devices is due to the presence of interface defects at the CVD-grown Gr/MoS$_2$ interface [6]. The polycrystalline nature of CVD-grown graphene and the transferring process of such CVD-graphene onto SiO$_2$/Si substrates could induce surface contamination and interface defect formation between CVD-Gr and MoS$_2$ [11,12].

Extending the performance limit of GMM devices is crucial for realizing Gr-based electronics. In this Letter, we report our fabrication of an exfoliated-graphene/MoS$_2$/Ti vertical GMM heterostructure via mechanical exfoliation and dry transfer of Gr and MoS$_2$ layers. The mechanical exfoliation technique affords a high-quality graphene layer and interface between Gr and MoS$_2$. The interface between Gr and MoS$_2$ exhibits near-ideal Schottky-diode behavior, thereby facilitating strong current rectification. Furthermore, in our study, we demonstrate a large current modulation exceeding $10^5$ together with a large ON state current density of ~$10^4$ A/cm$^2$. These values indicate superior device performance when compared with those of all other existing Gr-based vertical field effect transistor (FET) devices.

The schematic of our Gr/MoS$_2$/Ti device is shown in Fig. 1(a). First, a single-layer Gr bottom electrode was fabricated on a 300-nm-thick SiO$_2$/$n$-Si(001) substrate via mechanical exfoliation of Kish graphite. Second, multi-layer MoS$_2$ was exfoliated from a synthetic MoS$_2$ crystal (2D semiconductors Inc.) and deposited on the Gr electrode through the dry transfer technique [13,14]. The advantage of this transfer process is that we could prepare freshly cleaved Gr and MoS$_2$ surfaces, which were subsequently brought into contact with each other, thereby ensuring minimum contamination at the Gr/MoS$_2$ interface. Finally, Au (30 nm)/Ti (50 nm) metal top electrodes were fabricated on MoS$_2$ via standard electron beam (EB) lithography and EB evaporation. The cross-sectional transmission electron microscope (TEM) image of the device is shown in Fig. 1(b). Clear and distinct $MoS_2$ layers are observed, which indicates good device quality. We remark that the Gr layer is not visible in the TEM image because single-layer Gr cannot survive irradiation by high-energy electron beams. We fabricated a series of GMM devices with different $MoS_2$ thicknesses ranging from 2.4 to 24 nm and confirmed these thickness values using AFM measurements. These thicknesses correspond to the layer number range of $N$ = 4–37, assuming the single-monolayer thickness of $MoS_2$ to be 0.65 nm. For characterizing vertical transport across the GMM devices, we applied a source-drain bias $V_B$ across the Au/Ti electrode and the Gr layer. A back-gate bias $V_G$ was applied between the Si substrate and the Gr layer, thereby enabling control over the carrier concentration and thus the Fermi level of Gr. The Dirac point of the Gr layer for our series of devices was located in the range of $V_G$ = 0–15 V, and the carrier mobility for Gr was 3000–5000 cm$^2$/Vs, as determined from two-terminal resistance measurements. The junction areas for the series of devices were 1–3 μm$^2$.

First, we examine the transport properties of the GMM device for $N$ = 37. The photograph of this device is shown in Fig. 1(c). The current–voltage characteristics (*I*–*V*) were measured between electrodes A and B (shown in Fig. 1(c)) over a voltage range of $V_G$ = -50 to +50 V at 300K; the results are shown in Fig. 1(d). We note that the Dirac point of Gr is located at $V_G$ ~ +5 V. The current values are normalized with the junction area of 1 μm$^2$ for this device. Here, positive $V_B$ values correspond to electron flow from Gr to $MoS_2$. We observed a significant change in the *I*–$V_G$ curve with respect to change in $V_G$. In particular, the modulation of current is large in the positive-$V_B$ region, whereas it is relatively small in the negative-$V_B$ region. The $V_G$ dependence of the current density at $V_B$ = +0.5 V is shown in Fig. 1(e). A large current ON-OFF ratio of $I(V_G = +50\text{ V})/I(V_G = -50\text{ V})$ exceeding $10^5$ at $V_B = +0.5$ V was observed in our exfoliated-Gr/MoS$_2$/Ti vertical heterostructure. This value is significantly higher than the previously reported values observed in CVD-Gr/MoS$_2$/Ti devices, although the device structures of both devices are fairly similar [6].

In this section, we discuss the obtained *I–V* characteristics. In the positive-$V_B$ region, electron flow is directed from Gr to MoS$_2$. The conductance of the GMM device is dominated by the presence of the Schottky barrier at the Gr/MoS$_2$ interface. This Schottky barrier height is modulated by $V_G$ as can be inferred from Figs. 1(f) and (g), thereby enabling gate modulation of the conductance. On the other hand, in the negative-$V_B$ region, the conductance is dominated by the Schottky barrier at the Ti/MoS$_2$ interface; this barrier height does not change with $V_G$, and thus, there is very little gate modulation. This can be also seen from the asymmetry of the *I–V* curve in Fig. 1(d). At $V_G = -50$ V, the Fermi level of Gr assumes a minimal value, and consequently, the Schottky barrier height at the Gr/MoS$_2$ interface is maximum, as can be observed in Fig. 1(g). The *I–V* curve at $V_G = -50$ V exhibits strong current rectification characteristics, and further, from the asymmetry of the *I–V* curve, we note that this rectification is consistent with that of the Schottky barrier between Gr and MoS$_2$ [6]. On the other hand, at $V_G = +50$ V, the asymmetry of the *I–V* curve is reversed such that the current density at $V_B = +0.5$ V is larger than that at -0.5 V. In this case, the Fermi level of Gr is at its maximal value and the Schottky barrier at the Gr/MoS$_2$ interface is very small, as can be observed in Fig. 1(f). The conductance in this case is dominated by the MoS$_2$/Ti Schottky barrier; thus, the asymmetry is opposite to that for the Gr/MoS$_2$ interface. From the in-plane transport measurements on the MoS$_2$ monolayer with the Ti contact, we determined the Schottky barrier height between MoS$_2$ and Ti to be 0.13 eV.

Figures 2(a-d) show the ON-OFF characteristics for different MoS$_2$ layer numbers: $N$ = 14, 18, 25, and 37. The *I–V* curves at $V_G$ = +50 and -50 V are shown for each case. With increase in the MoS$_2$ layer number, the OFF current (current density at $V_G$ = -50 V) in the positive-$V_B$ region significantly decreases. The *I–V* curve at $V_G$ = -50 V becomes increasingly asymmetric for larger $N$ values, and thus, a larger degree of rectification is achieved. From the slope of the *I–V* curve in the negative-$V_B$ region at $V_G$ = -50 V, we determine the ideality factor $n$ of the Schottky barrier at the Gr/MoS$_2$ interface. In the thermionic emission model for the Schottky barrier, the current is expressed as;

$$J_{TE} = A^* T^2 \exp\left(-\frac{e\varphi_B}{k_B T}\right)\left\{\exp\left(\frac{eV_B}{nk_B T}\right) - 1\right\}, \tag{1}$$

where $A^*$ represents the effective Richardson constant, $\varphi_B$ the potential barrier height at the Gr/MoS$_2$ interface, $e$ the elementary charge, $k_B$ the Boltzmann constant, and $T$ the temperature [15]. The fitting results are indicated by solid lines in Figs. 2(a-d). The ideality factor is determined for all MoS$_2$ layer thickness values and plotted in Fig. 2(e). The ideality factor is close to unity when the MoS$_2$ layer number $N$ > 15, thereby suggesting near-ideal Schottky-diode behavior for these $N$ values. Such an ideal Schottky diode behavior has not been reported in previous experiments [6,10]. By carefully controlling the fabrication of the van der Waals interface between exfoliated-Gr and MoS$_2$ using mechanical exfoliation and the dry transfer technique, we obtained a significantly improved ideality factor in our device. Here, we remark that for lesser numbers of MoS$_2$ layers, the ideality factor significantly increases, and this increase suggests the contribution of non-Schottky-type conduction phenomena such as carrier tunneling through the MoS$_2$ layers.

The data for the ON current density $I_{ON}$ (current density at $V_B$ = +0.5 V and $V_G$ = +50 V) and OFF current density $I_{OFF}$ (current density at $V_B$ = +0.5 V and $V_G$ = -50 V) are shown in Fig. 3(a).

The ON current density is nearly constant for all MoS$_2$ layer thicknesses. We note that this ON current density is comparable to previously reported values in similar GMM devices [6]. Based on the ON and OFF density current data in Fig. 3(a), the ON-OFF ratio $I_{ON}/I_{OFF}$ data are plotted in Fig. 3(b). For comparison, the ON-OFF ratio $I_{ON}/I_{OFF}$ data determined in ref. 6 for the CVD-Gr/MoS$_2$/Ti device are also shown. Here, the ON-OFF ratios are defined at $V_B$ = +0.5 V for both sets of results for the purpose of comparison under identical bias conditions. The data in both cases exhibit a systematic change in the ON-OFF ratio with change in the MoS$_2$ layer number $N$. However, we find a significantly different layer-thickness dependence in our results. In particular, the ON-OFF ratio is significantly higher in our exfoliated-Gr/MoS$_2$/Ti device. We obtained a large ON current density of ~$10^4$ A/cm$^2$ along with a large current ON-OFF ratio using our GMM vertical FET. Since the ON current density is comparable between our results and those in ref. 6, we think that the large ON-OFF ratio in our GMM device is due to the significant reduction in the OFF current density as a consequence of the high quality of the exfoliated-Gr/MoS$_2$ Schottky interface.

Thus far, several studies have focused on the ON-OFF ratio and ON current density of various Gr-based FETs. A large ON-OFF ratio of $10^5$–$10^7$ has been achieved in Gr/WS$_2$/Gr vertical tunnel transistors and Gr/Si barristors [5,9]. However, in these devices, the ON current density tends to be small, lying in the range of $10^1$–$10^2$ A/cm$^2$. The Gr/MoS$_2$/metal vertical FET structure has been used to achieve an ON current density of $10^3$ A/cm$^2$ [6]. However, the maximum ON-OFF ratio is limited to ~$10^3$ in the range of $V_G$ = +60 and -80 V for 300-nm-thick SiO$_2$ gate dielectric layers. The maximum expected ON-OFF ratio for Gr/MoS$_2$/metal vertical FET can be roughly estimated using the thermionic emission model. From Eq. (1), we have $I_{ON}/I_{OFF}$ ~ exp[{$\mu_G(V_G$ = +50 V) - $\mu_G(V_G$ = -50 V)}/$k_B T$]. Thus, the ON-OFF ratio only depends on the modulation of graphene's Fermi level, which is given by $\Delta\mu_G = \mu_G(V_G = +50$ V$) - \mu_G(V_G = -50$ V$)$; thus larger $\Delta\mu_G$ values correspond to larger ON-OFF ratios. This value reaches ~0.4 eV for a SiO$_2$ gate dielectric thickness of 300 nm [16]. Therefore, at 300 K, the maximum expected ON-OFF ratio for our vertical FET in the range of $V_G = +50$ and $-50$ V is given by the relation $\exp(0.4$ eV$/k_B T) \sim 10^6$. We note that the ON-OFF ratios shown in Fig. 3(b) are already close to this limit. In fact, the ratio is limited not by the modulation of the Fermi level in our GMM structure but by the series resistance of the device; this series resistance includes the resistance of the Au/Ti electrode, contract resistance at the Ti/Gr interface, and the resistance of Gr. We estimated the total series resistance of our device to be about 2–3 k$\Omega$ at $V_G = +50$ V; this corresponds to a maximum ON current density of the order of $10^4$ A/cm$^2$ at $V_B = +0.5$ V. Therefore, either reducing the junction area or improving the carrier mobility of Gr (by fabricating the device on an h-BN substrate) will drive the ON-OFF ratio close to the limit. The achievement of both a large ON-OFF ratio (~$10^5$) and a large ON-current density in the range of ~$10^4$ A/cm$^2$ in an exfoliated-graphene/MoS$_2$/Ti vertical heterostructure demonstrates superior device performance when compared with those of other graphene-based vertical transistors. We believe that the use of this type of 2D crystal heterostructure can lead to further developments in electronics applications.

**Acknowledgements**

We are grateful to the Foundation for Promotion of Material Science and Technology of Japan (MST) for TEM analysis. This work was partly supported by Grants-in-Aid for Scientific Research from the Japan Society for the Promotion of Science (JSPS); the Science of Atomic Layers, a Grant-in-Aid for Scientific Research on Innovative Areas from the Ministry of Education, Culture, Sports and Technology (MEXT); and the Project for Developing Innovation Systems of the MEXT. S.Morikawa acknowledges the JSPS Research Fellowship for Young Scientists.

**Figure captions**

Fig. 1

(a) Schematic of Gr/MoS$_2$/metal (GMM) vertical heterostructure and measurement circuit. (b) Cross-sectional TEM image of the MoS$_2$ layer. (c) Optical microscope image of the GMM device. (d) Current–voltage ($I$–$V$) characteristics of the GMM device measured at different $V_G$ values at 300 K. (e) $V_G$ dependence of current $I$ through the device at $V_B$ = +0.5 V. (f,g) Schematic of band alignment of GMM device at (f) $V_G$ = +50 V and (g) $V_G$ = -50 V.

Fig.2

(a-d) The $I$–$V$ curve in the ON state ($V_G$ = +50 V, solid line) and OFF state ($V_G$ = -50 V, dashed line) for GMM device samples with different MoS$_2$ layer numbers $N$. The fitting results for the OFF state $I$-$V$ curves obtained using the thermionic emission theory for determining the ideality factor $n$ are also plotted. (e) Variation in the ideality factor $n$ for different values of $N$.

Fig.3

(a) The MoS$_2$ layer number ($N$) dependence of the ON current density $I_{ON}$ (current density at $V_B$ = +0.5 V and $V_G$ = +50 V) and OFF current density $I_{OFF}$ (current density at $V_B$ = +0.5 V and $V_G$ = -50 V). (b) The $N$ dependence of the current ON-OFF ratio $I_{ON}/I_{OFF}$. The $I_{ON}/I_{OFF}$ data reported in ref. 6 are also plotted. The overlapping data points ($N$ = 37) have been horizontally shifted for convenience of comparison. The broken lines are visual guides.

Figure 1

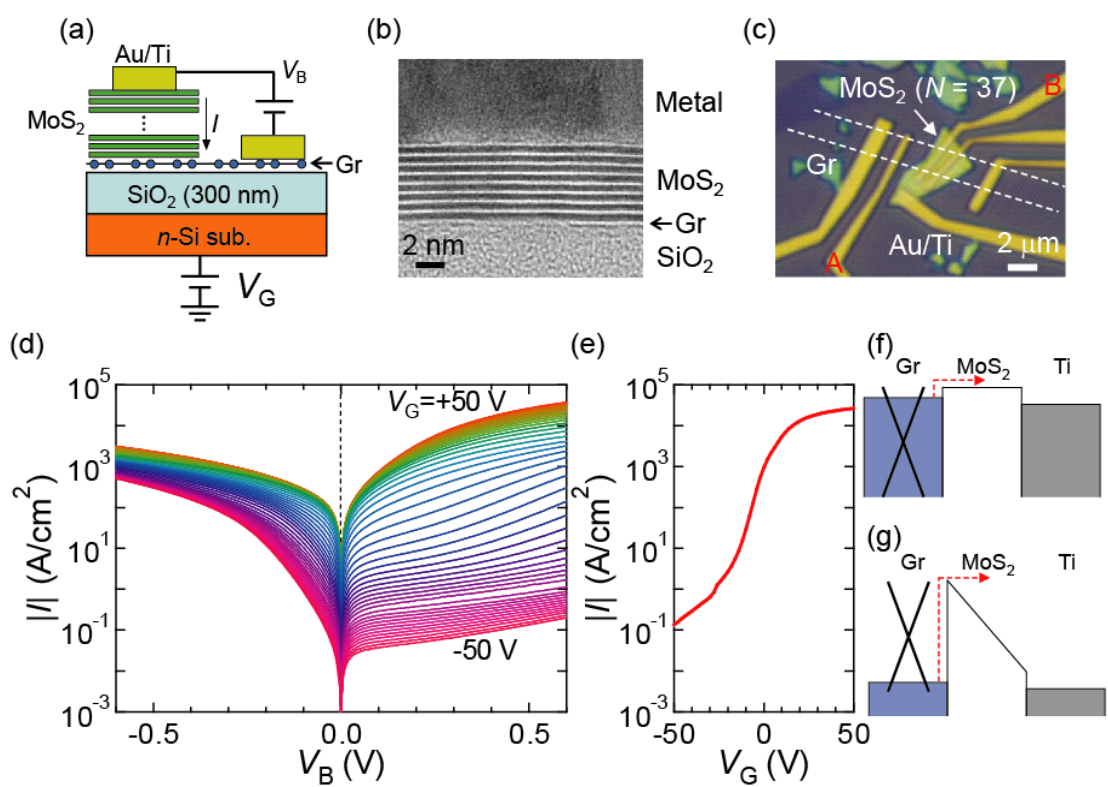

Figure 2

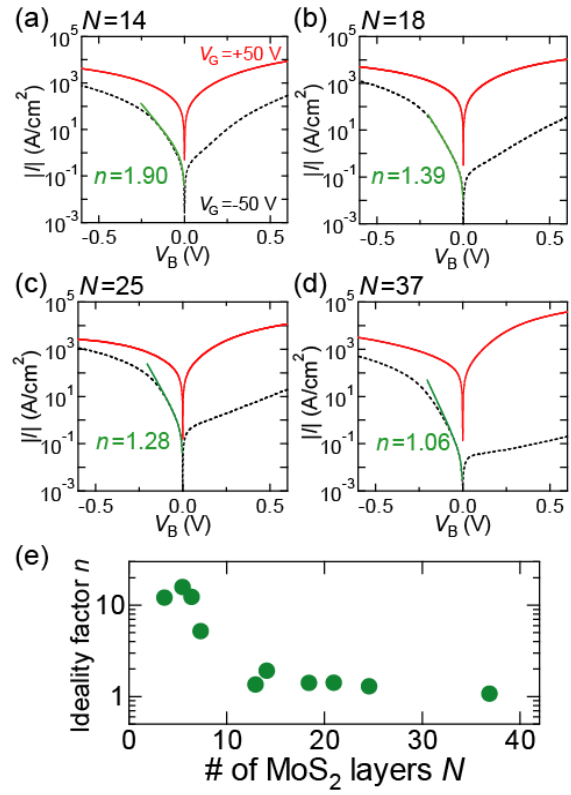

Figure 3

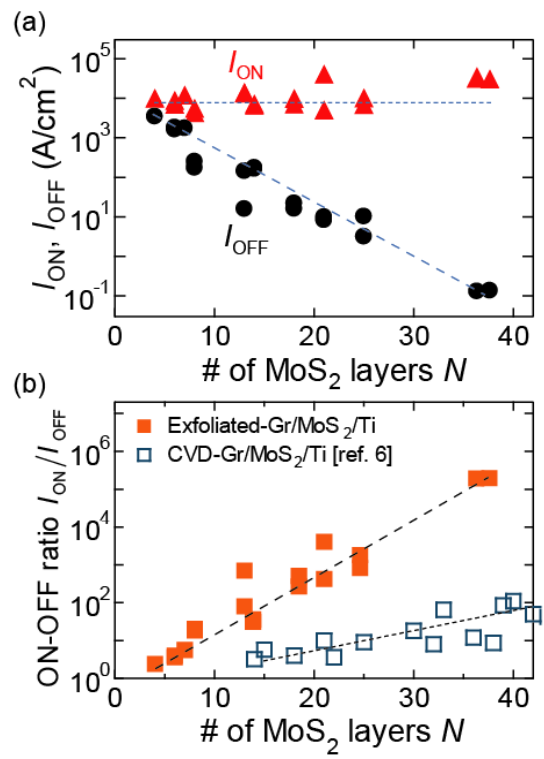